
\input phyzzx
\pubnum{$\caps 92/49$}
\titlepage
\title{A \ \ Note \ \ on \ \ the \ \ Green - Schwarz \ \ Mechanism \ \ in \ \
Open - String \ \ Theories}
\author{Augusto \ \ Sagnotti}
\address{\it Dipartimento \ di \ Fisica\break
Universit\`a di Roma \ ``Tor Vergata''\break
I.N.F.N. \ -- \ Sezione di Roma \ ``Tor Vergata''\break
Via \ della \ Ricerca \ Scientifica , \ 1\break
00133 \  Roma \ \ ITALY}
\vskip .6 truein
\abstract
{\baselineskip=14pt
An interesting feature of some open superstring models in $D < 10$ is the
simultaneous presence, in the spectrum, of gauge
fields and of a number of antisymmetric tensor fields.  In these cases the
Green-Schwarz mechanism can (and does) take a generalized form, resulting from
the combined action of all the antisymmetric tensors.  These novelties
are illustrated referring to some simple rational models in six dimensions, and
some
of their implications for the low-energy effective field theory are pointed
out.
}
\endpage
\pagenumber=2
\chapter{Introduction}
\vskip 24pt
Whereas the Green-Schwarz mechanism was originally discovered in the type-I
superstring theory in ten dimensions \Ref\gs{M.B. Green
and J.H. Schwarz, {\sl  Phys. Lett.} {\bf 149B} (1984) 117; \nextline
{\sl Nucl. Phys.} {\bf B255} (1985) 93.}, the
construction of the heterotic string \Ref\het{D.J. Gross, J.A. Harvey,
E. Martinec and R. Rohm, {\sl Nucl. Phys.} {\bf B256} (1985) 253.} soon
led to a widespread interest in its applications to models of oriented closed
strings.  These studies uncovered the deep relation that, in this case, holds
between
the cancellation of anomalies and the geometric property of modular invariance.
A fair, if somewhat crude, way to summarize these findings is by observing
that,
in theories of oriented closed strings, modular invariance removes the
ultraviolet region altogether, thus disposing of all anomalies at once.

The results obtained for models of
oriented closed strings, however, have no direct bearing
on the nature of the corresponding phenomenon in open-string
theories.   To wit, in this latter case modular invariance is not a property
of all amplitudes.  Rather, the consistency of open-string models results from
delicate cancellations taking place between some of the amplitudes.
In particular, the analysis of the ten-dimensional gauge anomaly of
ref. [\gs]~ involves three types of contributions (fig. 1). The first two
describe, respectively, the emission of six vectors from one of the boundaries
of the annulus and from the single boundary of the M\"obius strip.  Thus, they
determine the total irreducible contribution to the anomaly polynomial,
proportional in this case to $\rm Tr( F^6 )$.  As pointed out in ref. [\gs]~,
this contribution cancels precisely if the gauge group is $SO(32)$.  The nature
of the cancellation was nicely illustrated in ref. \Ref\pc{J. Polchinski and Y.
Cai, {\sl Nucl. Phys.} {\bf B296} (1988) 91.}: if $n$, the order of the $SO$
group, differs from $32$, there exists a spurious mode that, though projected
out of the closed string spectrum via the Klein-bottle diagram, couples to the
vacuum channel of both annulus and M\"obius diagrams.  The presence of
non-vanishing irreducible parts of the anomaly polynomial may then be linked to
a vacuum expectation value of this mode at ``genus-one-half'', resulting
from the contributions of the disk and the projective plane, by arguments
similar to those presented in ref. \Ref\bsnig{ M. Bianchi and A.
Sagnotti, {\sl Phys. Lett.} {\bf 211B} (1988) 407.}.
It should be appreciated that genus-one vacuum amplitudes capture the
whole essence of the  phenomenon, since the irreducible parts of the anomalies
draw their
origin from the region of coalescence of all vertices, where only the limiting
behavior
with respect to the surface moduli matters.  This is true for ultraviolet
divergences
as well, both in the $SO(32)$ superstring \Ref\gsdiv{M.B. Green and
J.H. Schwarz, {\sl Phys. Lett.} {\bf 151B} (1985) 21.} and in the $SO(8192)$
bosonic string \Ref\sobig{M.J. Douglas and B. Grinstein, {\sl Phys. Lett.} {\bf
183B} (1987) 52;\nextline S. Weinberg, {\sl Phys. Lett.} {\bf 187B} (1988)
287;\nextline
N. Marcus and A. Sagnotti, {\sl Phys. Lett.} {\bf 188B} (1987) 58.}.

Out of the three diagrams in fig. 1, the third (the ``non-planar''
diagram) is actually the most important one, since it is the home of the
Green-Schwarz mechanism.  Indeed, being regulated by the  momentum flow along
the tube, it does not contribute to the anomaly, a result that, in the limiting
field theory, may be ascribed [\gs] to a cancellation induced by new couplings
of the antisymmetric tensor.

The purpose of the present note is to illustrate some peculiar features of the
anomaly cancellation mechanism in open-string theories.  To this end, we
shall refer to one class of six-dimensional models introduced in refs.
\Ref\bss{M. Bianchi
and A. Sagnotti, {\sl Phys. Lett.} {\bf 247B} (1990) 517.} and
\Ref\bst{M.  Bianchi
and A. Sagnotti, {\sl  Nucl. Phys.} {\bf B361} (1991) 519.} where, as we shall
see, the
Green-Schwarz mechanism is at work in its full-fledged form.
This class
of open-string models is itself rather unconventional, the only available
derivation being [\bss] [\bst]~, as far as I know, one that starts from the
``parent'' closed string, as suggested in ref. \Ref\car{A. Sagnotti, {\it in}
Cargese '87, ``Non-Perturbative Quantum Field Theory'', \nextline
eds. G. Mack et al.
(Pergamon, New York, 1988), p. 521.}.  Thus, our original motivation was to
test to
a finer degree the consistency of these models. The novelties may
be related to the existence of a number of independent sectors of the spectrum,
and to the corresponding presence of a number of antisymmetric tensors
interacting with gauge fields [\bss]~[\bst]~. The very presence of
(anti)self-dual tensors coming from Ramond-Ramond sectors, noted in refs.
[\bss]~[\bst]~, is already a distinctive mark with respect to heterotic
models.  Even more interesting, as we shall see, is their role in the
anomaly cancellation mechanism.
\vskip 30pt
\chapter{The Generalized Green-Schwarz Mechanism}
\vskip 24pt
The nature of the problem is well illustrated by referring to the first class
of
rational open-string models of ref. [\bst]~, whose notation we adopt.  These
models
are chiral and supersymmetric in a six-dimensional space time.  In
this case a net total of $four$ self-dual antisymmetric tensors survive the
Klein bottle projection\foot{The sector $( 1 )$ contributes one antiself-dual
tensor, while each of the sectors $( 5 )$, $( \tilde{1} )$,
$\tilde{( 6 )}$, $\tilde{( 7 )}$ and $( \tilde{8} )$ contributes one self-dual
tensor.}of the closed string. Moreover, the corresponding open strings may have
sixteeen different sectors.  Their sixteen charge sectors are
constrained by the six tadpole conditions
$$ \eqalign{ &\sum_{i=1}^{8} n_i \ = \
\sum_{i=1}^{8} \tilde{n}_i \ = \ 16 \qquad ; \cr
&n_5 - n_1 + \tilde{n}_1 + \tilde{n}_6 + \tilde{n}_7 + \tilde{n}_8 \ = \ 8
\qquad ; \cr
&n_6 - n_2 + \tilde{n}_2 + \tilde{n}_5 + \tilde{n}_7 + \tilde{n}_8 \ = \ 8
\qquad ; \cr
&n_7 - n_3 + \tilde{n}_3 + \tilde{n}_5 + \tilde{n}_6 + \tilde{n}_8 \ = \ 8
\qquad ;\cr
&n_8 - n_4 + \tilde{n}_4 + \tilde{n}_5 + \tilde{n}_6 + \tilde{n}_7 \ = \ 8
\qquad . \cr}
\eqn\tadpole
$$
Still, they can accommodate
non-trivial symplectic gauge groups, such as $USp(8)^{ \otimes 4}$, with chiral
fermions in the representations $(\bf{8},\bf{8},\bf{1},\bf{1})$,
$(\bf{8},\bf{1},\bf{8},\bf{1})$, $(\bf{1},\bf{8},\bf{1},\bf{8})$ and
$(\bf{1},\bf{1},\bf{8},\bf{8})$.  In addition to the scalar multiplets
containing
these fermions, the
massless spectrum includes the supergravity multiplet, five tensor
multiplets and sixteen scalar multiplets from the closed sector, as well as
the gauge multiplet from the open sector.  For instance, to build a model with
a $USp(8)^{
\otimes 4}$ gauge group one may choose
$n_1 = n_2 = \tilde{n}_7 = \tilde{n}_8 = 8$,
thus correcting a typographical error in ref. [\bst]~.
This class of models therefore describes the interactions of a number
of antisymmetric tensors with non-trival gauge fields and chiral matter,
precisely what we are after.

In order to proceed further, we would like to construct the
anomaly polynomial for this case.  Since all fermions are in fundamental or
symmetric tensor representations of $Usp$ gauge groups, the polynomial
may be reduced to a standard form using only a few simple trace identities.
First of all, for fermions in the adjoint representation one finds
$$
\rm{Tr} F^2 \ = \ ( n + 2 ) \ \rm{tr} F^2
\eqn\adjquad
$$
and
$$
\rm{Tr} F^4 \ = \ ( n + 8 ) \ \rm{tr} F^4 \ + \ 3 \ ( \rm{tr} F^2 )^2
\eqn\qdjquar
$$
where, as usual, $\rm{tr}$ denotes the trace in the fundamental representation,
and where the two-form $F = {1 \over 2} F_{\mu \nu} dx^{\mu} dx^{\nu}$.
On the other hand, for fermions in the $(\bf{m},\bf{n})$ representation of
$\rm{G_1} \otimes \rm{G_2}$ one finds
$$
\rm{Tr}_{(\bf{m},\bf{n})} F^2 \ = \ \ m \ \rm{tr}_{(\bf n)} F^2 \ + \
n \ \rm{tr}_{(\bf m)} F^2
\eqn\fundquad
$$
and
$$
\rm{Tr}_{(\bf{m},\bf{n})} F^4 \ = \ m \ \rm{tr}_{(\bf n)} F^4 \ + \
n \ \rm{tr}_{(\bf m)} F^4 \ + \ 6 \ \rm{tr}_{(\bf m)} F^2 \ \rm{tr}_{(\bf n)}
F^2  \qquad .
\eqn\fundquar
$$

Let us begin by restricting our attention to a model with four types of quantum
numbers.  In this case, after imposing the tadpole
conditions, the anomaly polynomial is
$$
\eqalign{\rm{A} \ &= \ {1 \over 8} \ \Big\lbrace \ ( \rm{tr} {F_1}^2
)^2 \ + \ ( \rm{tr} {F_2}^2 )^2 \ + \ ( \rm{tr} {F_{\tilde{7}}}^2 )^2 \ + \ (
\rm{tr} {F_{\tilde{8}}}^2 )^2 \ \Big\rbrace \cr
&+ \ {1 \over 16} \ \Big\lbrace \ \rm{tr} {F_1}^2 \ + \
\rm{tr} {F_2}^2 \ + \ \rm{tr} {F_{\tilde{7}}}^2 \ + \ \rm{tr} {F_{\tilde{8}}}^2
\Big\rbrace \ \rm{tr} R^2 \cr
&- {1 \over 4} \Big\lbrace \ \rm{tr} {F_1}^2 \ \rm{tr} {F_{\tilde{7}}}^2 \ + \
\rm{tr} {F_1}^2 \ \rm{tr} {F_{\tilde{8}}}^2 \ + \ \rm{tr} {F_2}^2 \ \rm{tr}
{F_{\tilde{7}}}^2 \ + \
\rm{tr} {F_2}^2 \ \rm{tr} {F_{\tilde{8}}}^2 \ \Big\rbrace \cr
&- \ {1 \over 32} \ ( \rm{tr} R^2 )^2 \ \qquad \qquad , \cr }
\eqn\polysym
$$
where the two-form $R^{ab} = {1 \over 2} {R_{\mu \nu}}^{ab} dx^{\mu} dx^{\nu}$.

This result seems rather disappointing at first sight, since the residual
anomaly polynomial {\it does not factorize}.  Thus, the prospects for this
model (and for the whole construction of refs. [\bss]~[\bst]~) would appear
rather
bleak, since the conventional
Green-Schwarz mechanism does not apply in this case.  On the other hand,
the model contains a number of
antisymmetric tensors, and one may wonder whether they could realize a
more general type of Green-Schwarz mechanism by acting in a combined fashion.
The answer to this question may be deduced from the anomaly
polynomial, provided one regards it as a quadratic form in the field traces
and turns it into its diagonal form.  This is rather simple to do in this
case, and the result is
$$
\eqalign{\rm{A} \ &= \ - \ {1 \over 32} \ {\Big\lbrace \
\rm{tr} {F_1}^2 \ + \ \rm{tr} {F_2}^2 \ + \  \rm{tr} {F_{\tilde{7}}}^2 \ + \
\rm{tr} {F_{\tilde{8}}}^2 \ - \ \rm{tr} R^2 \ \Big\rbrace}^2 \cr
&+ \ {3 \over 32} \ {\Big\lbrace \
\rm{tr} {F_1}^2 \ + \ \rm{tr} {F_2}^2 \ - \  \rm{tr} {F_{\tilde{7}}}^2 \ - \
\rm{tr} {F_{\tilde{8}}}^2 \  \Big\rbrace}^2 \cr
&+ \ {1 \over 32} \ {\Big\lbrace \
\rm{tr} {F_1}^2 \ - \ \rm{tr} {F_2}^2 \ + \
\rm{tr} {F_{\tilde{7}}}^2 \ - \ \rm{tr} {F_{\tilde{8}}}^2 \
 \Big\rbrace}^2 \cr
&+ \ {1 \over 32} \ {\Big\lbrace \
\rm{tr} {F_1}^2 \ - \ \rm{tr} {F_2}^2 \ - \
\rm{tr} {F_{\tilde{7}}}^2 \ + \ \rm{tr} {F_{\tilde{8}}}^2 \
 \Big\rbrace}^2 \qquad . \cr
 }
\eqn\diag
$$

This expression displays the expected phenomenon: disposing of the anomaly
requires the combined action of
a number of antisymmetric tensors.  It should be appreciated that eq. \diag~
contains precisely six contributions, as many as the antisymmetric tensors in
the
model.  Moreover, the antiself-dual tensor belonging to
the supergravity multiplet is the only one that
couples to the gravitational Chern-Simons form, in
analogy with the standard case. Since all other tensor couplings involve
only combinations of
Yang-Mills Chern-Simons forms, one may investigate their nature
using the low-energy supergravity.  We shall return to this issue in the
next section.

Eq. \diag~ actually suggests the proper way to extend the result to models
with sixteen charge sectors.  The key observations are as
follows.  First of all, the gravitational Chern-Simons
form should couple only to the antisymmetric tensor in the supergravity
multiplet.  Moreover, the antisymmetric tensors that
are supposed to be at work in the cancellation mechanism belong to
the sectors contributing the massless tadpoles.  Since the $S$ matrix
of the conformal theory determines the vacuum channel,
one should be able to read from it all the proper combinations of Yang-Mills
Chern-Simons forms corresponding to the non-planar diagrams. Thus, the proper
generalizations of the last three lines in eq. \diag~should contain
combinations
of field traces weighted according to suitable lines of the $S$ matrix. This is
actually the case for the model with sixteen charge sectors, where one may
verify that the diagonal form of the anomaly polynomial is
$$
\eqalign{\rm{A} \ &= \ - \ {1 \over 2} \ {\Big\lbrace \
\sum_m S_{1m} \  \rm{tr} F_m^2 \ - \ 4 \ \rm{tr} R^2 \ \Big\rbrace}^2 \cr
&+ \ {1 \over 2} \sum_k \ {\Big\lbrace \
\sum_m S_{km} \ \rm{tr} F_m^2 \ \Big\rbrace}^2  \qquad ,  \cr}
\eqn\gen
$$
where $m$ runs over the range $( 1 - 8 )$ and $( \tilde{1} - \tilde{8} )$,
and where $k$ runs over the ``tadpole'' sectors $( 5 )$, $( \tilde{1} )$,
$( \tilde{6} )$, $( \tilde{7} )$ and $( \tilde{8} )$.

Defining polynomials $F^{2(i)}$ corresponding to the various lines of
eq. \gen~, one may write
$$
{\rm A} \ = \ - \ {1 \over 2} \sum\limits_{ij} \ \eta_{ij} \ F^{2(i)} \
F^{2(j)}
\qquad ,
\eqn\short
$$
where $\eta$ is the Minkowski metric with signature $(1-n)$. If, following
standard
practice, the
eight-form in eq. \short~is converted into the Green-Schwarz counterterm
$$
\Delta L \ = \ + \ {1 \over 2} \sum\limits_{ij} \ \eta_{ij} \ F^{(i)} B_{(j)}
\qquad ,
\eqn\gscount
$$
the modified field strengths for the antisymmetric tensors are
$$
H_{(i)} \ = \ d B_{(i)} \ + \ \omega_{(i)} \qquad ,
\eqn\chern
$$
where $\omega_{(i)}$ denote the combinations of (Yang-Mills and gravitational)
Chern-Simons forms
corresponding to the various field traces in eq. \gen~.  We have repeated the
analysis for several of the models in ref. [\bst]~, reaching identical
conclusions.  Namely, in general the Green-Schwarz mechanism is the result of
the
combined action of several antisymmetric tensors. In four-dimensional models,
one may
envisage interesting applications of this generalized mechanism to models with
a number of anomalous $U(1)$ factors in their gauge groups.

It should be appreciated that the structure of the residual anomaly
polynomial of eq. \gscount~ is suggestive of an
$SO(1,5)$ symmetry relating the antisymmetric tensors, broken only by the
explicit form of
the Chern-Simons couplings. In the next section we shall see that, in general,
an $SO(1,n)$ symmetry of this kind is precisely in the spirit of the limiting
supergravity
theory for this class of models.
\vskip 30pt
\chapter{Field Equations of
Six-Dimensional N = 2b Supergravity Coupled to Vector and Tensor Multiplets}
\vskip 24pt
We would like to construct the field equations of a class of
supergravity models related to the string spectra of the preceding section. The
restriction to field equations is natural \Ref\ms{N. Marcus and J.H. Schwarz,
{\sl Phys. Lett.} {\bf 115B} (1982) 111. } \Ref\jhs{J.H. Schwarz, {\sl
Nucl. Phys.}
{\bf B226} (1983) 289.} in all cases where (anti)self-dual
tensors are present. These models describe the coupling
of chiral $N = 2b$ supergravity in six dimensions to a number of tensor
multiplets, as well as to vector multiplets.  This extends the work of
ref. \Ref\rom{L.J. Romans, {\sl Nucl. Phys.} {\bf  B276} (1986) 71.}, where the
coupling to a number of tensor multiplets was discussed, and the work of ref.
\Ref\ns{H. Nishino and E. Sezgin, {\sl Nucl. Phys.} {\bf B278} (1986) 353.},
where the
coupling to a single tensor multiplet and to arbitrary matter was constructed
\foot{In this case the antiself-dual tensor in the supergravity multiplet may
be combined with the self-dual tensor in the tensor multiplet to give an
ordinary antisymmetric tensor with no (anti)self-duality, and one may write
an action using standard techniques.}.  We adopt notation and
spinor conventions of ref. [\ns]~. As in refs. [\jhs]~and [\rom]~, we confine
our attention
to terms of lowest order in the fermions.

Let us begin by recalling the results of ref. [\rom], while rephrasing them in
a
slightly different fashion.  In addition to the supergravity multiplet, that
contains a graviton, a left-handed gravitino and an antiself-dual tensor, we
consider $n$ tensor multiplets, each containing a self-dual tensor, a
right-handed spinor and a scalar.  Following ref. [\rom]~, we let the $n$
scalar
fields parametrize the coset space $SO(1,n)/SO(n)$, and we introduce the
$SO(1,n)$ matrix
$$
V \ = \ \pmatrix{v_0&v_M\cr {x^m}_0&{x^m}_M \cr}		\qquad .
\eqn\scalars
$$
Since $V$ is a (pseudo)orthogonal matrix, its elements satisfy the relations
$(r = 0, ... ,M)$
$$
\eqalign{&{\tilde{v}}^r \ v_s \ + \ {{\tilde{x}}^r}_{\ m} \ {x^m}_s \ = \
{\delta^r}_s \cr
&v_r \ {\tilde{v}}^r \ = \ 1 \cr
&{x^m}_r \ {{\tilde{x}}^r}_{\ n} \ = \ {\delta^m}_n   \qquad , \cr}
\eqn\son
$$
where ${\tilde{v}}^r = {\eta}^{rs} v_s$ and ${{\tilde{x}}^r}_m = -
{\eta}^{rs} {x^m}_s$, with $\eta$ the Minkowski metric with signature
$(1 - n)$.
{}From these expressions one may derive the composite $SO(n)$ connection
$$
{S_{\mu}}^{[mn]} \ = \ ( \ \partial_{\mu} \ {x^m}_r \ ) \ {{\tilde{x}}^r}_{\ n}
\qquad ,
\eqn\conn
$$
antisymmetric in $(m,n)$ because of eq. \son~.  The scalar kinetic term is
then built out of
$$
{P^m}_{\mu} \ = \ \sqrt{{1 \over 2}} \ ( \ \partial_{\mu} \ v_r \ ) \
{{\tilde{x}}^r}_{\ m}	\qquad ,
\eqn\kin
$$
where $P$ satisfies $D_{[{\mu}}  {P^m}_{{\nu}]} \ = \ 0$.

The $(n + 1)$ tensor fields ${A^r}_{\mu \nu}$ are then taken to transform in
the
fundamental representation of $SO(1,n)$, and out of them one defines the
composite field strengths
$$
\eqalign{ H_{\mu \nu \rho} \ &= \ v_r \ {F^r}_{\mu \nu \rho} \cr
{K^m}_{\mu \nu \rho} \ &= \ {x^m}_r \ {F^r}_{\mu \nu \rho}\qquad . \cr}
\eqn\ten
$$
To lowest order in the spinor fields, the tensor equations are then defined to
be the conditions that $H$ be self-dual and that $K^m$ be antiself-dual:
$$
\eqalign{ H_{\mu \nu \rho} \ &= \ {\tilde{H}}_{\mu \nu \rho} \cr
{K^m}_{\mu \nu \rho} \ &= \ - \ {{\tilde{K}}^m}_{\ \ \mu \nu \rho} \qquad . \cr
}
\eqn\dual
$$
These equations and the Bianchi identities then imply the second-order
equations
$$
\eqalign{ D_{\mu} \ H^{\mu \nu \rho}  \ &= \ - \ \sqrt{2} \ {P^m}_{\mu} \
K^{m \mu \nu \rho} \cr
D_{\mu} \ K^{m \mu \nu \rho} \ &= \ - \ \sqrt{2} \ {P^m}_{\mu} \
H^{\mu \nu \rho} \qquad . \cr}
\eqn\second
$$

For convenience, all Weyl fermions satisfy an $Sp(2)$ Majorana
condition.  A number of identities familiar from the four-dimensional case
then apply, for instance [\ns]
$$
\bar{\chi} {\gamma}^{r_1 ... r_n}  \lambda \ = \ (-1)^n \
\bar{\lambda} {\gamma}^{r_n ... r_1} \chi  \qquad .
\eqn\flip
$$
With all these ingredients one may then show that, to lowest order in the
fermions, the field equations of the spinor fields
$$
\eqalign{ &{\gamma}^{\mu \nu \rho}  D_{\nu} \psi_{\rho} \ + \
H^{\mu \nu \rho} \ \gamma_{\nu} \psi_{\rho} \ - \ { i \over 2} \
K^{m \mu \nu \rho} \ \gamma_{\nu \rho} \chi^m \ - \
{i \over \sqrt{2}} \ {P^m}_{\nu} \ \gamma^{\nu} \gamma^{\mu}
 \chi^m \ = \ 0 \cr
&\gamma^{\mu} D_{\mu} \chi^m \ - \ {1 \over 12} \ H_{\mu \nu \rho} \
\gamma^{\mu \nu \rho} \chi^m \ - \ {i \over 2} \ K^{m \mu \nu \rho} \
\gamma_{\mu \nu}  \psi_{\rho} \ + \ {i \over \sqrt{2}} \
{P^m}_{\nu} \ {\gamma}^{\mu}  {\gamma}^{\nu} \psi_{\mu} \ = \ 0 \cr}
\eqn\spin
$$
transform into
$$
\eqalign{&R_{\mu \nu} \ - \ {1 \over 2} g_{\mu \nu}  R \ - \  H_{\mu \rho
\sigma}
{H_{\nu}}^{\rho \sigma} \ - \ {K^m}_{\mu \rho \sigma}
{{K^m}_{\nu}}^{\rho \sigma} \ - \ 2 {P^m}_{\mu} {P^m}_{\nu} \ + \
g_{\mu \nu} \ {P^m}_{\rho} P^{m \rho} \ = \ 0 \cr
&D_{\mu} P^{m \mu} \ - \ {\sqrt{2} \over 3} \  H^{\mu \nu \rho}
 {K^m}_{\mu \nu \rho} \ =
\ 0 \qquad \qquad ,
 \cr}
\eqn\bose
 $$
as well as into eqs \dual~, under the local supersymmetry transformations
$$
\eqalign{ &\delta  {e_{\mu}}^m \ = \ - \ i \ \bar{\epsilon}  \gamma^m
 \psi_{\mu} \cr
&\delta \psi_{\mu} \ = \ D_{\mu} \ \epsilon \ + \ {1 \over 4}
\ H_{\mu \nu \rho} \ \gamma^{\nu \rho}  \epsilon \cr
&\delta {A^r}_{\mu \nu} \ = \ i \ {\tilde{v}}^r \ {\bar{\psi}}_{[ \mu}
\gamma_{ \nu ]}  \epsilon \ - \ {1 \over 2} \ {{\tilde{x}}^r}_m \
{\bar{\chi}}^m  \gamma_{\mu \nu} \epsilon \cr
&\delta \chi^m \ = \ - \ {i \over \sqrt{2}} \ {\gamma}^{\mu} {P^m}_{\mu} \
\epsilon \ + \ {i \over 12} \ {K_{\mu \nu \rho}}^m \ {\gamma}^{\mu \nu \rho}
\epsilon \cr
&\delta v_r \ = \ {x^m}_r \ \bar{\epsilon} {\chi}^m \qquad \qquad . \cr}
\eqn\susy
$$
The extension to higher orders is then tedious but possible in principle, for
instance using superspace techniques,
as in ref. \Ref\hw{P.S. Howe and P.C. West, {\sl Nucl. Phys.} {\bf B238}
(1984) 181.}.  In particular, if the field strengths
in the fermionic transformations are extended to their supercovariant forms,
one may verify that the supersymmetry algebra closes on the bosons in terms of
all local symmetries.

Our next step will be extending the construction to the case when a number
of vector multiplets are coupled to the model.  To this end, we begin by
including
in the tensor field strengths suitable Chern-Simons
forms\Ref\manton{E. Bergshoeff, M. de Roo, B. de Wit and P. van
Nieuwenhuizen,\nextline
{\sl Nucl. Phys.} {\bf B195} (1982) 97;\nextline
G.F. Chapline and N.S. Manton, {\sl Phys. Lett.} {\bf 120B} (1983) 105.}.
Let us recall that, under a gauge transformation, the Chern-Simons form
$$
\omega \ = \ {\rm tr} ( \ A \ dA \ - {{2 i g} \over 3} \ A^3 \ )
\eqn\chern
$$
varies according to
$$
\delta \omega \ = \ d {\rm tr} ( \ \Lambda \ dA \ )	\qquad .
\eqn\chgauge
$$
If one modifies the tensor field strengths according to
$$
F^r \ = \ d A^r \ - \ c^{rz} \ \omega_z		\qquad ,
\eqn\strength
$$
where $z$ labels the various factors of the gauge group, and where $c^{rz}$ is
a
matrix of constants, related to the elements of the $S$ matrix of the conformal
theory,
the invariance of $F^r$ under the vector gauge transformations demands that
$$
\delta A^r \ = \ c^{rz} \ {\rm tr}_z ( \ \Lambda \ dA \ )		\qquad .
\eqn\gauge
$$
The Bianchi identity is also modified, and eqs. \second~ become
$$
\eqalign{ D_{\mu} \ H^{\mu \nu \rho}  \ &= \ - \ \sqrt{2} \ {P^m}_{\mu} \
K^{m \mu \nu \rho} \ - \ {1 \over {8e}} \ \epsilon^{\mu \nu \rho \alpha \beta
\gamma}
\ v_r \ c^{rz} \ {\rm tr}_z ( F_{\mu \alpha} \ F_{\beta \gamma} )
\cr
D_{\mu} \ K^{m \mu \nu \rho} \ &= \ - \ \sqrt{2} \ {P^m}_{\mu} \
H^{\mu \nu \rho} \ + \ {1 \over {8e}} \ \epsilon^{\mu \nu \rho \alpha \beta
\gamma}
\ {x^m}_r \ c^{rz} \ {\rm tr}_z ( F_{\mu \alpha} \ F_{\beta \gamma} ) \qquad ,
\cr}
\eqn\second
$$
where $e$ denotes the determinant of the vielbein.

The field equations of the spinor fields now include additional terms,
$$
\eqalign{ &{\gamma}^{\mu \nu \rho}  D_{\nu} \psi_{\rho} \ + \
H^{\mu \nu \rho} \ \gamma_{\nu} \psi_{\rho} \ - \ { i \over 2} \
K^{m \mu \nu \rho} \ \gamma_{\nu \rho} \chi^m \ - \
{i \over \sqrt{2}} \ {P^m}_{\nu} \ \gamma^{\nu} \gamma^{\mu}
 \chi^m  \cr
&- \  {1 \over {2 \sqrt{2}}} \ \gamma^{\sigma \tau} \ {\gamma^{\mu}} \
v_r \ c^{rz} \ {\rm tr}_z ( F^{\sigma \tau} \ \lambda ) \ = \ 0 \cr
&\gamma^{\mu} D_{\mu} \chi^m \ - \ {1 \over 12} \ H_{\mu \nu \rho} \
\gamma^{\mu \nu \rho} \chi^m \ - \ {i \over 2} \ K^{m \mu \nu \rho} \
\gamma_{\mu \nu}  \psi_{\rho} \ + \ {i \over \sqrt{2}} \
{P^m}_{\nu} \ {\gamma}^{\mu}  {\gamma}^{\nu} \psi_{\mu} \cr
&- \ {i \over {2 \sqrt{2}}} \ {x^m}_r \ c^{rz} \ {\rm tr}_z ( \gamma^{\mu \nu}
\lambda
\ F_{\mu \nu} )
\ = \ 0 \qquad . \cr}
\eqn\dirac
$$
Moreover, the Einstein equation becomes
$$
\eqalign{R_{\mu \nu} \ &- \ {1 \over 2} g_{\mu \nu}  R \ - \  H_{\mu \rho
\sigma}
{H_{\nu}}^{\rho \sigma} \ - \ {K^m}_{\mu \rho \sigma}
{{K^m}_{\nu}}^{\rho \sigma} \ - \ 2 {P^m}_{\mu} {P^m}_{\nu} \cr &+  \
g_{\mu \nu} \ {P^m}_{\rho} P^{m \rho} \ + \
2 v_r \ c^{rz} \ {\rm tr}_z ( F_{\lambda \mu} {F^{\lambda}}_{\nu} \ - \
{1 \over 4} g_{\mu \nu} F^2 ) \ = \ 0  \qquad , \cr}
\eqn\einstein
$$
while the equation of the scalar fields becomes
$$
D_{\mu}  P^{m \mu} \ - \ {\sqrt{2} \over 3} \  H^{\mu \nu \rho}
 {K^m}_{\mu \nu \rho} \ + \ {1 \over {2 \sqrt{2}}} \ {x^m}_r c^{rz} \
{\rm tr}_z ( F_{\alpha \beta} F^{\alpha \beta} ) \ = \ 0 \qquad .
\eqn\scalv
$$
To these we should add the supersymmetry transformations for the vector
multiplets,
$$
\eqalign{&\delta \lambda \ = \ - \ {1 \over {2 \sqrt{2}}} \ F_{\mu \nu} \
{\gamma}^{\mu \nu} \epsilon \cr
&\delta A_{\mu} \ = \ - \ {i \over {\sqrt{2}}} ( \bar{\epsilon} \gamma_{\mu}
\lambda )
\qquad \qquad , \cr}
\eqn\susyv
$$
and the corresponding field equations
$$
\eqalign{ ( v_r c^{rz}) \ \gamma^{\mu} D_{\mu} \lambda \ &+ \ {1 \over
{\sqrt{2}}}
{P^m}_{\mu} ({x^m}_r c^{rz}) \ \gamma^{\mu} \lambda \ + \
{1 \over {2 \sqrt{2}}} \ ( v_r c^{rz}) \ F_{\lambda \tau} \gamma^{\mu}
\gamma^{\lambda
\tau} \psi_{\mu} \cr &+ \ {i \over {2 \sqrt{2}}} \ ({x^m}_r c^{rz}) \
\gamma^{\mu \nu}
\chi^m \ F_{\mu \nu} \ = \ 0  \cr}
\eqn\gaugini
$$
and
$$
\eqalign{ ( v_r c^{rz}) \ D^{\mu} F_{\mu \nu} \ &+ \ \sqrt{2} \ ({x^m}_r
c^{rz}) \
P^{m \mu} F_{\mu \nu}
- \ ( v_r c^{rz}) \ F_{\rho \sigma} {H_{\nu}}^{\rho \sigma} \cr &- \
({x^m}_r c^{rz}) \ F^{\rho \sigma} {K^m}_{\nu \rho \sigma} \ = \ 0 \qquad .
\cr}
\eqn\yang
$$
Finally, the supersymmetry transformation of the antisymmetric tensors acquires
an
additional contribution,
$$
\delta {A^r}_{\mu \nu} \ = \ - \ c^{rz} \ {\rm {tr}_z} ( A_{[ \mu} \ \delta
	A_{\nu ]} ) \qquad ,
\eqn\extrat
$$
necessary in order that the commutator algebra give rise to the proper vector
gauge transformation.  Again, with proper supercovariantizations, the
supersymmetry
algebra on the bosons closes on all local symmetries.

It should be appreciated that the vector couplings change under
$SO(1,n)$ transformations, since the $A^r$ tensors are coupled to different
combinations
of Chern-Simons forms.  As a result, from eqs. \gaugini ~ and \yang ~one may
see that,
for instance, the scalar fields should be restricted to the region where
$$
v_r \ c^{rz} \ > \ 0		\qquad \qquad ,
\eqn\manif
$$
since at the boundaries of this region the coupling constants of the vector
fields
become infinite.  These restrictions have in fact a number of conventional
analogues in supergravity models. For instance, with a single tensor multiplet,
and in the absence
of vector couplings, one might conclude that the only physical scalar should
live on the
hyperbola
$$
( v_0 )^2 \ - \ ( v_1 )^2 \ = \ 1 \qquad \qquad .
\eqn\onlyone
$$
On the other hand when, according to common practice, the scalar matrix is
parametrized
in terms of the ``dilaton'' field $\phi$ [\ns]~, writing
$$
V \ = \ \pmatrix{{{\rm cosh}( \sqrt{2} \phi )}&{{\rm sinh}( \sqrt{2} \phi )}\cr
{{\rm sinh}( \sqrt{2} \phi )}&{{\rm cosh}( \sqrt{2} \phi )} \cr}
\qquad , \eqn\onescal
$$
one is implicitly covering only one branch of the curve. Moving to the other
branch
would require that one continue the dilaton according to $\sqrt{2} \phi
\rightarrow
\sqrt{2} \phi  +  i \pi$ but, in open-string theories, this would alter the
sign of the
vector kinetic term, that comes from ``genus-one-half''.

\vskip 24pt
It is a pleasure to thank I. Antoniadis,
C. Bachas, M. Bianchi, S. Ferrara, P. Fr\'e, C. Kounnas and G. Pradisi for
several stimulating discussions.  Part of this work was carried out while the
author was visiting the Theory Group of CERN, the Physics Department of UCLA
and
the Laboratoire de Physique Th\'eorique of Ecole Polytechnique.  The kind
hospitality of these Institutions is gratefully acknowledged.  This work was
supported in part by E.C. contract $SCI-0394-C$.
\vskip 24pt
\centerline{\bf Figure Captions}
\vskip 24pt
{\bf Figure 1}

Contributions to the gauge anomaly in open-string theories:
(a) planar diagram ; (b) non-orientable diagram ; (c) non planar diagram.

\endpage
\refout
\end